\documentclass[12pt, twoside]{article}

\usepackage{a4wide}
\usepackage{epsfig}  
\usepackage{cite}

\newcommand{\Extra}{False}
\newcommand{\extra}[2]{ \ifthenelse{\equal{\Extra}{True}}{
 #1}{[[[#2]]]}  }

\newcommand{\bq}{\begin{equation}}
\newcommand{\eq}{\end{equation}}
\newcommand{\bqa}{\begin{eqnarray}}
\newcommand{\eqa}{\end{eqnarray}}

\newcommand{\eps}{\varepsilon_0 \varepsilon_r}
\newcommand{\dif}{\ensuremath{\textrm{d}}}

\newcommand{\micro}{\ensuremath{\mu}}
\newcommand{\degrees}{\ensuremath{^\circ}}

\author{H.J.J. Verheijen\footnote{Also at: Eindhoven University of
	   Technology} \/ and
	   M.W.J. Prins\footnote{email: prins@natlab.research.philips.com}\\  \\
\small
\centerline{\emph{Philips Research Laboratories Eindhoven, Prof. Holstlaan 4,}}\\
\small
\centerline{\emph{5656 AA  Eindhoven, The Netherlands}}}

\title{Reversible electrowetting and trapping of 
          charge: model and experiments}

\begin{document}

\maketitle

\abstract{
We derive a model for voltage-induced wetting, so-called electrowetting,
from the principle of virtual displacement.
Our model includes the possibility that charge is trapped in or on the 
wetted surface.
Experimentally, we show reversible electrowetting for an aqueous 
droplet on an insulating layer of 10~\micro m thickness. 
The insulator is coated with a highly fluorinated layer impregnated with oil,
providing a contact-angle hysteresis lower than 2\degrees.
Analyzing the data with our model, we find that until a threshold voltage of
$240$~V, the induced charge remains in the liquid and is not trapped.
For potentials beyond the threshold, 
the wetting force and the contact angle saturate, in 
line with the occurrence of trapping of charge in or on the insulating layer.
The data are independent of the polarity of the applied electric field, 
and of the ion type and molarity. 
We suggest possible microscopic origins for charge trapping.}

\section*{Introduction}
\begin{figure}[htb]
\centerline{\epsfig{file=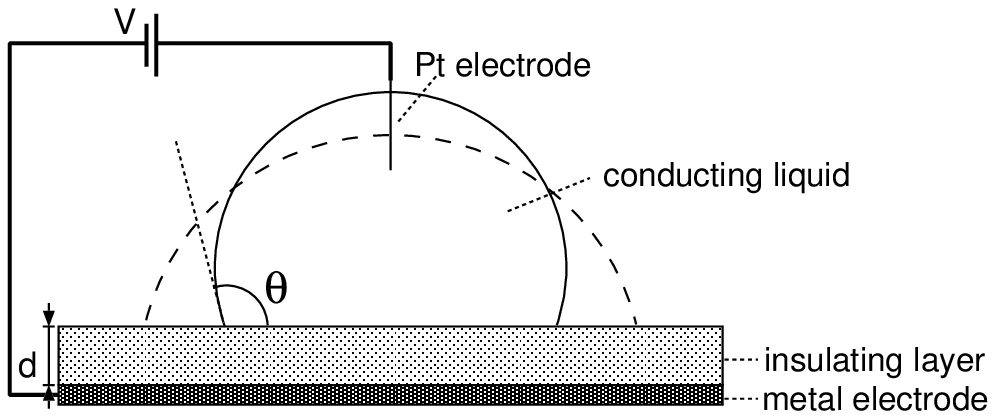,width=7cm}}
~\\
 \caption{Schematic drawing of an 
 electrowetting experiment. A droplet of a 
 conducting liquid is placed on an insulating layer of thickness $d$,
 which is deposited on a metal counter electrode. 
 Application of a potential $V$ between the droplet and the metal electrode
 changes the free energy of the droplet and results
 in a decrease of the contact angle $\theta$. The resulting droplet shape is
 indicated by the dashed line.
  }
\label{fig:lang_droplet}
\end{figure}

It is possible to reduce the contact angle of a droplet on a surface by 
applying an electric field between the 
conducting liquid and a counter electrode underneath the liquid~\cite{Minnema,
Beni0, Welters, Sondag93, Berge95, Whitesides} as shown in 
Fig.~\ref{fig:lang_droplet}.
This so-called electrowetting effect was observed first by 
Minnema~\cite{Minnema} in 1980 using 
an insulator between liquid and counter electrode and by Beni~\cite{Beni0} in 
1981 with the liquid directly on the counter electrode.
The electric field results in a distribution of charge that changes
the free energy of the droplet, causing the droplet to
spread and wet the surface.
In systems with the liquid in direct contact with the solid 
electrode, the potential drops across a diffuse ionic double layer at the 
interface.
In systems with an insulating layer of several micrometers thickness
between the solid electrode and the 
liquid~\cite{Minnema,Welters,Berge95}, the main voltage drop 
appears across the insulating layer.

In this paper, we use an insulating layer between the counter electrode and
the aqueous solution to enhance the electrowetting 
force~\cite{Berge95}, achieving reversible wetting by a suitable top coating.
Previously, limits have been observed for the voltage-induced reduction
of contact angle: at high electric fields, the contact angle 
saturates~\cite{Berge95, Welters}. 
We consider the possibility that
trapping of charge in or on the insulating layer affects the contact angle.
We define charge to be trapped when the charge is bonded more 
strongly to the insulating layer than to the liquid. 
First, we derive the theory of electrowetting from the principle of 
virtual displacement. 
This provides a flexible method to extend Young's equation to include
the influence of an arbitrary charge distribution.
We consider the case that a sheet of trapped charge 
is present in or on the insulating layer.
Next, we present a measurement of the contact angle as a function of applied
voltage and we extract the potential resulting from the trapped charge
as a function of applied voltage. 
Finally, we suggest possible microscopic origins for trapping of charge.

\section*{Electrowetting model}
\subsubsection*{Virtual displacement, no trapped charge.}
\begin{figure}[htb]
 \centerline{\epsfig{file=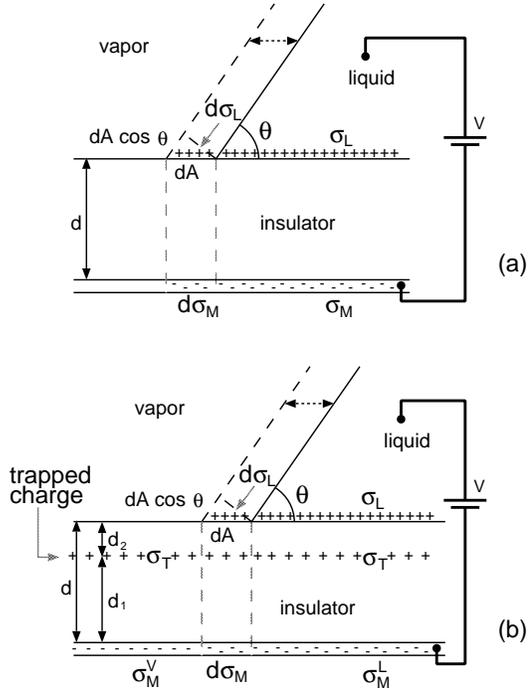,width=7cm}}
~\\
 \caption{(a) Schematic picture of the virtual displacement of the 
  contact line in the presence of a potential across the 
  insulator. An infinitesimal increase 
  in base area $\dif A$ at fixed voltage $V$ changes the free 
  energy of the droplet, as a result of a change in interface area and the
  placement of additional charge $\dif \sigma_L$ and image charge 
  $\dif \sigma_M$.
  (b) The virtual displacement of the contact line in the presence of a sheet of
  trapped charge. Now, the infinitesimal increase $\dif A$ alters the 
  free energy not only via the charge distribution between the electrode and 
  the liquid, but also via the charge distribution below the vapor phase.
     }
\label{fig:ew_virt}
\end{figure}

A droplet spreads until it has reached a minimum in free energy, determined by
cohesion forces in the liquid and adhesion between the liquid and the surface. 
In general, the energy required to create an interface is given 
by $\gamma$, the surface tension [N/m] or surface free energy [J/m$^{2}$]. 
In case of an applied potential, a change in the electric charge distribution 
at the liquid/solid interface changes the free energy.
We define our thermodynamic system as the droplet, the insulating layer,
the metal electrode and the voltage source. 
Throughout the entire derivation, we assume that the system is in equilibrium 
at constant potential $V$. 
We focus on the change in free energy due
to an infinitesimal increase in base area of the droplet on the solid
surface, surrounded by vapor. When a potential $V$ is applied, 
a charge density $\sigma_L$ builds up in the liquid phase 
and induces an image charge density $\sigma_M$ on the metal electrode.
Figure~\ref{fig:ew_virt}(a) shows the edge of a droplet and its virtual
displacement.
An infinitesimal increase of the base area $\dif A$ results in a 
contribution to the free energy from the surface energies as well as 
an energy contribution due to the additional charge density 
$\dif \sigma_L$ in the liquid and 
its image charge density $\dif \sigma_M$ on the metal electrode. 
The voltage source performs the work,
$\dif W_B$.
The free energy ($F$) of the system can be written in differential form:
\bq
  \dif F = \gamma_{SL}\dif A -\gamma_{SV}\dif A 
	  +\gamma_{LV}\dif A \cos \theta  +\dif U -\dif W_B \ ,
  \label{eq:E_common_no}
\eq
where $U$ is the energy required to create the electric field between the 
liquid and the counter electrode.
The parameters $\gamma_{SL}$, $\gamma_{SV}$ and $\gamma_{LV}$ are the 
free energies of the solid/liquid, solid/vapor and liquid/vapor 
interface respectively for the situation in the \emph{absence} of an 
electric field. The contact angle, $\theta$, is the angle
between the liquid/vapor interface and the solid/liquid interface
at the contact line (see Fig.~\ref{fig:lang_droplet})~\cite{dtheta}.
Mechanisms for energy dissipation, which may cause contact-angle hysteresis, 
are not taken into account.

Let us first consider the situation in the absence of an externally applied 
voltage, so $\dif U = \dif W_B = 0$. When $\dif F/\dif A = 0$, we find 
the minimum in free energy, relating the surface energies to the  
contact angle. This equation was obtained by Young~\cite{Young} in 1805:
\bq
   \gamma_{LV} \cos \theta = \gamma_{SV} -\gamma_{SL} \ .
   \label{eq:Young}
\eq

For a non-zero potential, we have to include the energy of the
charge distribution. In Fig.~\ref{fig:ew_virt}(a),
the droplet with charge density $\sigma_L$ is at constant voltage $V$, while 
the metal electrode with charge density $\sigma_M=-\sigma_L$ is at ground
potential.
The electrostatic energy per unit area below the liquid is given by:
\bq
  \frac{U}{A} = \int\limits^d_0 \frac{1}{2} \vec{E} \vec{D} \dif z \ ,
  \label{eq:energy-dens}
\eq
where $z$ is the coordinate perpendicular to the surface, 
$d$ the thickness of the insulating layer,
$\vec E$ the electric
field and $\vec{D}$ the charge displacement, with 
$\vec{D}=\eps  \vec{E}$~\cite{Jackson}.
The increase of free energy due to the charge distribution in the liquid, 
upon an infinitesimal increase of droplet base can be written as:
\bq
  \frac{\dif U}{\dif A}  =  \frac{1}{2} \ d \ E \ D  
          = \frac{1}{2} d \frac{V}{d} \sigma_L      
     = \frac{1}{2} V \sigma_L                  \ .
     \label{eq:UL_0}
\eq
The electric field originating from 
the liquid/vapor boundary of the droplet (the so-called fringing or stray 
field) makes a constant contribution to the free energy: 
this contribution remains
unaltered when the contact line is displaced by $\dif A$.
Therefore, the stray fields do not contribute to $\dif U$.
The voltage source performs the work to redistribute the charge;
per unit area the work is given by~\cite{Feynman}:
\bq
  \frac{\dif W_B}{\dif A} =  V \sigma_L  \ .
  \label{eq:WB_0}
\eq
Calculating the minimum of free energy, we get Young's equation 
with an additional electrowetting term $\gamma_{EW}$, the electrowetting 
force per unit length due to the applied potential:
\bq
 \gamma_{LV} \cos \theta  =  \gamma_{SV}-\gamma_{SL} + \gamma_{EW} \ ,
     \label{eq:F_ew_no}
\eq
with the electrowetting force:
\bq
         \gamma_{EW} =  \frac{1}{2}\frac{d}{\eps} \sigma_L^2 \ ,
   \label{eq:F_ew}
\eq
where we 
used $\sigma_L = \eps V/d$ (Gauss' law), with $\varepsilon_r$ the 
dielectric constant of the insulating layer and $\varepsilon_0$ the permittivity
of vacuum.
We can rewrite Eqs.(\ref{eq:F_ew_no}) and (\ref{eq:F_ew}) to get the well-known relation between 
$\theta$ and $V$ for electrowetting~\cite{Welters, Berge95, Sondag93}:
\bq
  \gamma_{LV} \left[ \cos \theta(V) - \cos \theta_0 \right]  =  
       	    \frac{1}{2}\frac{\eps}{d}V^2  \ ,
	    \label{eq:theta_EW}
\eq
where $\theta_0$ is the contact angle at zero volt.

\subsubsection*{Influence of charge trapping.}
When we apply a potential difference between the liquid and the metal
electrode, electric forces work on the ions in the liquid and pull 
them toward the insulating layer. 
There is a possibility that charge becomes trapped in or on the insulating
layer when the interaction of the ions with the solid is stronger 
than with the liquid.
In the three-phase region, ions are trapped when the de-trapping 
time is large compared to the typical vibration times of the contact line.
As a result of excitations of the droplet, 
e.g. thermal, mechanical or voltage-induced vibrations, 
a density of trapped charge arises on both sides of the contact line.

As yet we do not specify the precise nature of the trapped charge
(e.g. electronic or ionic); we simply assume that a layer of charge 
with constant surface charge density $\sigma_T$ is trapped in 
the insulating layer 
at distance $d_2$ from the top of the insulator as shown in 
Fig.~\ref{fig:ew_virt}(b).
To determine the change in electrostatic energy due to the infinitesimal 
base area increase [$\dif U$ in Eq.(\ref{eq:E_common_no})], we have to take 
into account the electrostatic contribution below the 
liquid $\dif U_{L}$, 
as well as the contribution below the vapor phase, $\dif U_V$:
\bq
  \dif U = \dif U_L -\dif U_V \ .
  \label{eq:E_common_yes}
\eq
The sign difference is due to the fact that the virtual
displacement $\dif A$ increases the solid/liquid interface while the solid/vapor
interface is decreased.
We assume that the trapped charge is distributed uniformly at
constant depth, extending sideways to the left of the contact line to a length 
scale of at least  the insulator thickness.
Then, the charge density at the liquid/vapor interface due to fringing fields
at the edge of the droplet is unaltered by the
virtual displacement. Therefore, we omit the electrostatic energy of the 
fringing field in Eq.(\ref{eq:E_common_yes}).

\begin{figure}[htb]
\centerline{\epsfig{file=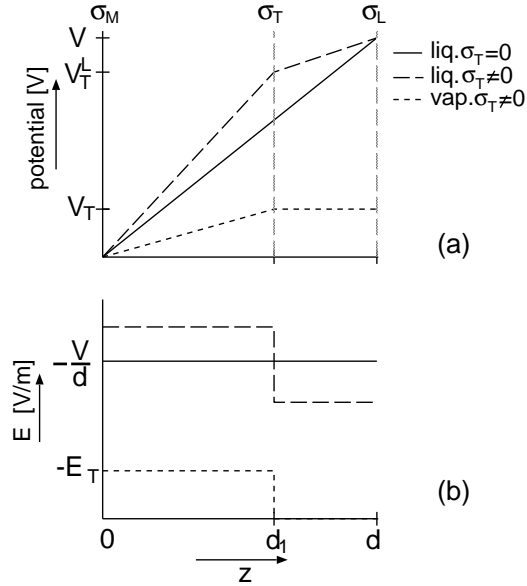,width=7cm}}
~\\
\caption{Sketch of the potential (a) and the electric field
(b) in the insulator beneath the liquid and the vapor phase. 
Below the liquid phase, the potential and
electrostatic field without trapped charge ($\sigma_T=0$, solid lines) and with
trapped charge ($\sigma_T \neq 0$, long dash) are shown. Below the vapor
phase, the curves in the presence of trapped charge are shown (short dash).
The voltage drop across the diffuse ionic double layer is neglected.
  }
\label{fig:pot_ew}
\end{figure}

The potential as a function of the depth in the insulator is sketched in
Fig.~\ref{fig:pot_ew}(a). 
The solid line shows the potential in the absence of trapped charge. 
For the case of charge trapping, the potential beneath the liquid phase is 
indicated by the long dashed line while the short dashed line shows the 
potential beneath the vapor phase.
The vertical line indicates the depth where the trapped charge is situated.
Figure~\ref{fig:pot_ew}(b)
shows a plot of the electrostatic fields, with $E=-\nabla V$.
It is 
clear that trapping of charge lowers the electric field at the liquid/solid
interface and should consequently reduce the electrowetting force.
The charge density of the trapped charge, $\sigma_T$, is at  
potential $V_T^L$ below the liquid and at $V_T$ below the vapor phase.
On the metal electrode below the liquid, the charge 
density is \mbox{$\sigma_M^L = -(\sigma_L+\sigma_T)$}.
The charge density on the metal electrode below the vapor phase 
is $\sigma_M^V=-\sigma_T$.
Using the general expression for the energy density 
[Eq.(\ref{eq:energy-dens})], we find the electrostatic energy density below the 
liquid phase:
\bqa
  \frac{\dif U_L}{\dif A} & = & \frac{1}{2} d_1 E_1 D_1 +
	     \frac{1}{2} d_2 E_2 D_2 \nonumber \\
     &=& \frac{1}{2} d_1 \frac{V_T^L}{d_1} (\sigma_T+\sigma_L) + 
	 \frac{1}{2} d_2 \frac{V-V_T^L}{d_2} {\sigma_L} \nonumber \\
     &=& \frac{1}{2} V_T^L \sigma_T + \frac{1}{2} V \sigma_L   \ .
     \label{eq:UL}
\eqa
The energy to create the charge distribution below the vapor phase equals:
\bq
  \frac{\dif U_V}{\dif A}  =  \frac{1}{2} d_1 \frac{V_T}{d_1} \sigma_T 
	   =  \frac{1}{2} V_T \sigma_T \ .
  \label{eq:UV}
\eq
The work performed by the voltage source per unit area is given by
Eq.(\ref{eq:WB_0}).
Using Gauss' law, we find the following relationships between the charge 
densities and the potentials:
\bqa
   \label{eq:surpot_a}  
      \sigma_T &=& \frac{\eps V_T}{d_1}  \\
   \label{eq:surpot_b}
     \sigma_L &=& \frac{\eps (V-V_T)}{d}  \\
    V_T^L & = & V_T + \frac{d_1}{d}(V-V_T) \ .
  \label{eq:surpot}
\eqa

Using Eq.(\ref{eq:E_common_no}), Eq.(\ref{eq:WB_0}),
Eqs.(\ref{eq:E_common_yes})--(\ref{eq:surpot}) and $\dif F / \dif A = 0$, 
we recover Eqs.(\ref{eq:F_ew_no}) and (\ref{eq:F_ew}), the modified Young
equation.
With Eq.(\ref{eq:surpot_b}), we find the following relation for the
contact angle modulation in the presence of trapped charge:
\bq
     \gamma_{LV} \left[\cos \theta(V) - \cos \theta_0 \right] =
	               \frac{1}{2}\frac{\eps}{d} (V-V_T)^2  \ .
   \label{eq:theta_trp}
\eq
The electrowetting force is proportional to the square of the 
applied voltage minus the voltage due to charge trapping. 
This causes a reduction of the electrowetting force.

\section*{Results}
We used a system as shown in Fig.~\ref{fig:lang_droplet}. 
On a silicon substrate, a conducting aluminum layer (100~nm) and an
insulating layer of parylene-N (10~\micro m, $\varepsilon_r=2.65$, 
chemical vapor deposited as in Ref.~\cite{Welters}) were applied. 
Subsequently a thin hydrophobic AF1600 top coating~\cite{AF1600} was deposited
by spincoating a 0.1~wt\% solution of AF1600 (DuPont) in FC726 (3M)  
at 1000 r.p.m. during 30~s, resulting in a layer of approximately 
30~nm thickness.
In order to abtain a low contact-angle hysteresis, we impregnated the coating with 
silicon oil.
The sample was left in silicon oil for several hours.
Before the experiment the excess of oil was removed; 
this is possible because the contact angle of silicon oil on AF1600
is about 40\degrees~\cite{Schneemilch}.
The in-liquid electrode was a platinum wire. 
The droplet consisted of 10~\micro l of an aqueous solution, 
with 1.0~M, 0.1~M, 0.01~M KCl or 0.1~M K$_2$SO$_4$.

\begin{figure}[htb]
  \centerline{\epsfig{file=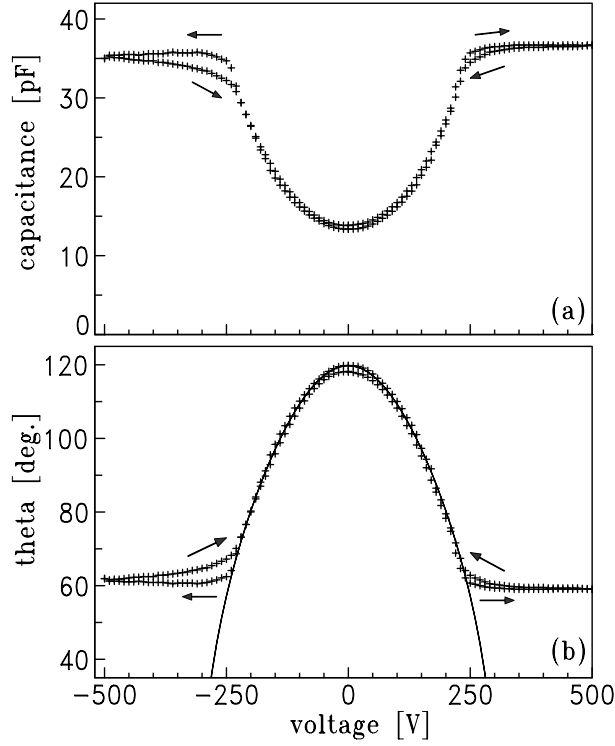,width=8cm}}
~\\
  \caption{(a) The capacitance between a 10~\micro l liquid droplet 
  and counter electrode as a function of applied dc-potential.
  The insulator thickness is 10~\micro m. We used a 700~Hz ac-signal with 
  5~V amplitude and a sweep rate $\sim10$~V/s. 
  (b) The contact angle derived from the capacitance measurement. The
  contact-angle hysteresis is less than 2\degrees \/ in the range $-240<V<240$~V.
  For higher voltages, the contact angle saturates around 60\degrees. The
  continuous line is according to Eq.(\ref{eq:theta_EW}), using
  $\theta_0=119$\degrees, $d=10$~\micro m, $\varepsilon_r=2.65$ and
  $\gamma_{LV}=72$~mN/m.
  }
\label{fig:vth_high}
\end{figure}

The capacitance was measured between the platinum electrode and the metal 
counter electrode as a function of applied voltage,
using a 700~Hz, 5~V ac-signal which was superimposed on the dc-voltage.
The capacitance gives a measure of the contact area between liquid and
surface and is shown in 
Fig.~\ref{fig:vth_high}(a). The potential of the liquid was increased to 500~V
and subsequently decreased to 0~V; the opposite voltage polarity was measured on a
different spot to avoid possible interference with previously trapped charge.
The time interval between measurement points (1~s) was more than an order
of magnitude longer than the time required for spreading of the droplet
(experimentally verified to be about 20~ms~\cite{Verheijen}). 
We find that the droplet base increases by nearly a factor of three 
due to the applied voltage. The
droplet recovers its original shape upon removal of the electric potential.
Measurements with the solutions of different molarities and ion types 
resulted in identical curves.

The contact angle $\theta$ was derived from the measured capacitance, using
the known dielectric constant and thickness of the insulating layer, 
the droplet volume and the droplet shape. This electrical measurement method 
gives the contact angle with an accuracy of 2\degrees; details are described 
in  Ref.~\cite{Verheijen}.
Figure~\ref{fig:vth_high}(b) shows the contact angle, derived from the 
capacitance measurement along with the theoretical curve according to
Eq.(\ref{eq:theta_EW}).
The contact-angle hysteresis is less than 2\degrees.
To our knowledge, such a high degree of reversibility 
has not yet been reported in electrowetting experiments.
We attribute the low value of contact-angle hysteresis
to the penetration of the oil into nano-pores of the amorphous 
fluoropolymer layer~\cite{AF1600}, which reduces the already very low
surface heterogeneity of the top coating (for water on non-impregnated
AF1600, the contact-angle hysteresis is about 7\degrees).
At zero volt, the value of the contact angle on the impregnated surface
agrees well with the advancing contact angle on the non-impregnated surface.
This indicates that the fluoropolymer determines the surface energy 
rather than the silicon oil.

We can distinguish two regions in the plot. In the region
$-240<V<240$~V, the measured contact angle is consistent with 
the theoretical
contact angle of Eq.(\ref{eq:theta_EW}) within 2\degrees.
According to our model [Eq.(\ref{eq:theta_trp})], 
this means that at low voltages the ionic charge remains
in the liquid without being trapped on or in the insulating layer. 
Within this voltage range we demonstrated reproducible droplet
spreading for more than 10$^5$ times.
At higher voltages, we notice contact-angle saturation at 
about 60\degrees and 
differences between the advancing and receding curves.
To investigate the possibility that the saturation of contact angle
is caused by trapping of charge, 
we applied a voltage
higher than 240~V to the droplet and while maintaining the applied voltage,
adsorbed the liquid into a tissue.
Afterward, when blowing humid air across the sample, we observed
a condensation pattern~\cite{Berge95} at the position and in the shape of the 
droplet base. 
During subsequent electrowetting experiments in this area, we 
detected a nonzero electrostatic potential, having
the sign of the previously applied potential, which points 
at the presence of charge.
Therefore, we attribute the condensation pattern to the preferential deposition
of polar water molecules on the charged area of the surface.
Finally, we noticed a decrease of the electrostatic surface potential after a 
large grounded droplet was placed on the charged coating; no decrease of the 
potential was observed when the large droplet was electrically floating. 
These experiments prove that contact-angle saturation is accompanied 
by charging of the insulating coating. The charge can be removed by an
electrical shortcut between the metal electrode and the surface of the 
insulator.

\begin{figure}[htb]
  \centerline{\epsfig{file=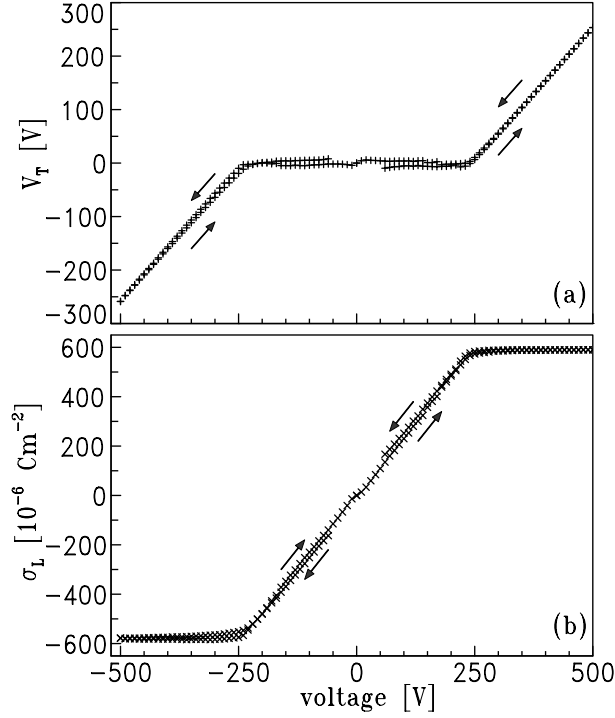,width=8cm}}
~\\
  \caption{(a) The voltage due to trapped charge, $V_T$, as a
  function of applied potential. $V_T$ is derived from the data of 
  Fig.~\ref{fig:vth_high} and Eq.(\ref{eq:theta_trp}). 
  For potentials below the threshold of $\pm240$~V, the 
  voltage due to trapped charge equals zero; 
  for higher potentials, charge gets trapped. 
(b) The surface charge density in the liquid, $\sigma_L$, calculated using
   Eq.(\ref{eq:surpot_b}). The charge density in 
  the liquid increases  until a threshold voltage is reached, beyond which 
  it saturates. Note that the charge density is
  of the same order as $10^{-4}$ monolayer of unit charge. 
      }
\label{fig:vtrapped}
\end{figure}

We have analyzed the data of Fig.~\ref{fig:vth_high} with our model for
voltage-induced wetting in the presence of trapped charge 
[Eq.(\ref{eq:theta_trp})].
We ascribe the difference between the measured contact angle and the
contact angle of the old model [Eq.(\ref{eq:theta_EW})] to the voltage of
trapped charge ($V_T$).
Figure~\ref{fig:vtrapped}(a) shows the resulting plot for $V_T$.
We notice a threshold voltage $V_{th}$ of $240\pm10$~V. For $|V|<V_{th}$, the
voltage of trapped charge equals zero and the electrowetting force
$\gamma_{EW}\propto V^2$. For $|V|>V_{th}$, $V_T \approx V-V_{th}$ and
$\gamma_{EW}\propto (V-V_T)^2$. This means that above the threshold voltage, 
almost all charge added by the voltage source gets trapped in or on the 
insulating layer.
Figure~\ref{fig:vth_high}(b) shows the charge density in the liquid phase
$\sigma_L$. For voltages below the threshold, $\sigma_L$ increases linearly with 
increasing applied voltage. Beyond the threshold voltage, $\sigma_L$ remains approximately
constant, in line with the saturation of the contact angle.

\section*{Discussion and Conclusions}
The data of the previous section show a threshold-like saturation behavior for
the electrowetting force and for the charge density in the liquid. The voltage
of trapped charge shows a linear increase beyond the threshold voltage. The
curves are symmetric for positive and negative 
potential, and independent of the ion type, ion valence and ion molarity 
that we have tested. Furthermore, the advancing curve is consistent with the 
receding curve, indicating that the trapped charge is released upon lowering of
the applied voltage.

Let us now consider possible microscopic origins for trapping of charge.
We defined trapped charge as charge which has a stronger interaction with the
insulating layer than with the liquid. 
Clearly, the underlying charge bonding mechanism cannot have a chemical nature,
as an expected dependence on voltage polarity, ion type, valence and molarity
was not observed.
Charge trapping could occur due to the attractive electrostatic force between 
ions in the liquid and the metal counter electrode. When the electrostatic force
on the ion exceeds the force between the liquid and the ion, it moves toward 
the insulating layer and remains in or at the insulating layer~\cite{Chudleigh}. 
The ion might exchange some of its hydration shell for a bond with the surface.
Although this model predicts a threshold-like trapping behavior, a
dependence on the valence of the ions is expected, in disagreement
with our experimental results.

At the threshold electric field, the electrowetting force is of the same order
of magnitude as or larger than the surface tensions in our system
(for $V=240$~V, $\gamma_{EW}= 68$~mN/m). Therefore, we propose that it 
is possible that instabilities in the liquid/solid interface or the liquid/vapor
interface occur. Small charged droplets or molecular clusters could move into 
nano-pores of the insulating layer and become trapped.
In this concept, no dependence on molarity, ion type, valence of the ions or
polarity of the applied field is expected. 
While this line of thought seems in
agreement with the behavior of $V_T$ (a threshold and subsequently a 
slope close to one), 
further research is needed to determine the microscopic mechanisms of charge
trapping. 

In conclusion, the principle of virtual displacement provides a transparent
method to calculate the influence of an arbitrary charge distribution
on the contact angle. 
The virtual change of electric energy is calculated by integrating 
the energy density of the electrostatic field.
Using this method, we derived a model for electrowetting that accounts for
a reduction of the electrowetting force by the assumption that charge is 
trapped in or on the insulating layer [Eq.(\ref{eq:theta_trp})].

We demonstrated reversible electrowetting using an aqueous droplet on 
a sample with a 10~\micro m thick parylene insulating layer and 
a highly fluorinated hydrophobic AF1600 top coating which was impregnated 
with silicon oil. 
The measured contact-angle hysteresis was below 2\degrees. 
For voltages between $-240$~V and $240$~V the charge remains in the liquid and
is not trapped in or on the insulating layer. At higher voltages, charge gets
trapped with a threshold-like behavior, limiting the charge density that can be
induced in the liquid. 
We observed that charge remains at areas where the droplet has receded in a 
high-voltage state and that discharging of the surface is possible with 
a zero-voltage droplet. For all solutions tested, the absolute value of the
threshold voltage is independent of the polarity of the applied voltage, the 
ion type, ion molarity and ion valence (Cl$^-$ vs. SO$_4^{2-}$).

From an experimental standpoint, the distribution of the trapped charge 
should be measured quantitatively with
for instance a scanning Kelvin probe. 
This will clarify the dependence of the trapped charge density on the applied 
voltage and on the distance with respect to the contact line.
The mechanisms of de-trapping of charge are interesting and need to be studied 
in more detail.
The dependence of the threshold electric field on the insulator
thickness, measured for different salts, solvents  and top coatings
may provide information on the mechanisms that cause the trapping of charge.
Eventually, when trapping of charge in the insulator can be avoided, it may
become possible to reversibly reach complete wetting of a surface by an 
applied electric potential.

\subsubsection*{Acknowledgement}
We thank W. Welters and A. Kemmeren for coating preparation and 
for valuable discussions.

%
% -- Bibliography 
%

%
% Figures
%
\clearpage
\section*{Figures}

~\\

~\\

\end{document}